\documentclass{emulateapj}

\shorttitle{Stellar Wind Torques}
\shortauthors{Matt \& Pudritz}

\begin{document}

\title{Accretion-Powered Stellar Winds II:\\Numerical Solutions for Stellar Wind Torques}




\author{Sean Matt\altaffilmark{1} and Ralph E. Pudritz\altaffilmark{2}}

\affil{$^1$Department of Astronomy, University of
Virginia, P.O. Box 400325, Charlottesville, VA 22904-4325; seanmatt@virginia.edu}

\affil{$^2$Physics and Astronomy Department, McMaster University,
Hamilton, ON L8S 4M1, Canada; pudritz@physics.mcmaster.ca}


\begin{abstract}

In order to explain the slow rotation observed in a large fraction of
accreting pre-main-sequence stars (CTTSs), we explore the role of
stellar winds in torquing down the stars.  For this mechanism to be
effective, the stellar winds need to have relatively high outflow
rates, and thus would likely be powered by the accretion process
itself.  Here, we use numerical magnetohydrodynamical simulations to
compute detailed 2-dimensional (axisymmetric) stellar wind solutions,
in order to determine the spin down torque on the star.  We discuss
wind driving mechanisms and then adopt a Parker-like (thermal pressure
driven) wind, modified by rotation, magnetic fields, and enhanced mass
loss rate (relative to the sun).  We explore a range of parameters
relevant for CTTSs, including variations in the stellar mass, radius,
spin rate, surface magnetic field strength, the mass loss rate, and
wind acceleration rate.  We also consider both dipole and quadrupole
magnetic field geometries.

Our simulations indicate that the stellar wind torque is of sufficient
magnitude to be important for spinning down a ``typical'' CTTS, for a
mass loss rate of $\sim 10^{-9} M_\odot$ yr$^{-1}$.  The winds are
wide-angle, self-collimated flows, as expected of magnetic rotator
winds with moderately fast rotation.  The cases with quadrupolar field
produce a much weaker torque than for a dipole with the same surface
field strength, demonstrating that magnetic geometry plays a
fundamental role in determining the torque.  Cases with varying wind
acceleration rate show much smaller variations in the torque
suggesting that the details of the wind driving are less important.
We use our computed results to fit a semi-analytic formula for the
effective Alfv\'en radius in the wind, as well as the torque.  This
allows for considerable predictive power, and is an improvement over
existing approximations.

\end{abstract}


\section{Introduction} \label{sec_intro}

For more than half a century, the spin rates and the angular momentum
evolution of stars have been topics of vigorous study.  We know that
stellar winds are responsible for the spinning down of late-type
(later than F2) main sequence stars \citep[][]{parker58,
schatzman62, kraft67, skumanich72, soderblom83, kawaler88,
macgregorbrenner91, barnessofia96, bouvier3ea97}.  There is still
progress to be made on main sequence star spins \citep{barnes03}, but
perhaps the largest open questions remain at the pre-main-sequence
phase, which determines the ``initial conditions'' for the spin
histories of stars.


By the time intermediate/low mass ($\la 2 M_\odot$) pre-main-sequence
stars become optically visible (T Tauri stars; TTSs), they already
have ages around $10^{5}$ -- $10^{6}$ yrs.  A large fraction of TTSs
(called classical TTSs; CTTSs) are observed to actively accrete
material from a disk at a rate within a wide range of $\sim 10^{-8}
M_\odot$ yr$^{-1}$ \citep[e.g.,][]{johnskrullgafford02}.  At this
rate, the angular momentum accreted from the orbiting disk should spin
up the stars to a substantial fraction of breakup speed in a short
amount of time (comparable to their ages).  The fact that the stars
are also still contracting \citep[e.g.,][]{rebullea02}, and that they
presumably were accreting at much higher rates before they became
optically visible, further adds to expectation of fast rotation.

Large data sets for the spins of TTSs in star formation regions and
clusters of different ages \citep[see][for a compilation]{rebull3ea04}
show that approximately half of the stars are rotating rapidly and do
seem to spin up as expected as they approach zero-age main sequence
\citep{vogelkuhi81, bouvier3ea97, rebull3ea04, herbstea07}.  However,
the surprise is that the other $\sim$half of TTSs exhibit much slower
rotation rates ($\sim 10$\% of breakup speed) at all ages.  Recent
studies have shown a correlation between slow rotation and the
presence of an a accretion disk \citep[see
  especially][]{ciezabaliber07}, though this idea has been
controversial in the past \citep[e.g.,][]{stassunea99, herbstea00,
  stassunea01, herbstea02}.  This is still an open issue, but it is
clear than an efficient angular momentum loss or regulation mechanism
is operating for the slow rotators.

Although alternative ideas have been proposed since \citep[][see
  \citealp{mattpudritz07coolstars} for a history]{konigl91, shuea94},
\citet{hartmannstauffer89} offered the first potential explanation for
the slow rotators, namely that massive stellar winds may be
responsible for carrying off substantial angular momentum \citep[see
  also][]{toutpringle92}.  In \citet[][hereafter Paper
  I]{mattpudritz05l}, we extended this idea to consider the effects of
the magnetic interaction between the star and disk, and we used a
1-dimensional scaling from the solar wind angular momentum loss to
estimate the torque for TTSs.  The scaling suggested that, for an
observationally constrained dipole magnetic field strength of 200 G
\citep[e.g.,][]{johnskrullea99, bouvierea07, johnskrull07iau,
  johnskrull07, smirnovea03, yang3ea07}, it might indeed be possible
for the stellar wind to extract enough angular momentum to explain the
slow rotators.  For stellar winds to balance the accreted angular
momentum, the wind outflow rate needs to be a substantial fraction of
the accretion rate.  In Paper I, we suggested that this is possible,
if a fraction of the energy liberated by the accretion process
actually powers the stellar wind.

The pre-main-sequence phase is, in fact, marked by powerful outflows
\citep{reipurthbally01}.  In the most powerful sources, due to the
large linear momenta of the outflows \citep{koniglpudritz00}, the
X-ray luminosities \citep{decampli81}, and possible detection of
rotation \citep{bacciottiea02, andersonea03, coffeyea04,
ferriera3ea06, coffeyea07}, it appears that most of the
flow arises from the accretion disk, rather than the star.  It is not
clear what fraction of the total outflow may actually originate from
the star, and thus how powerful are the stellar winds compared to
main-sequence phase winds or compared to their accretion rates.

There is some observational evidence for powerful stellar winds from
CTTSs, as distinguished from inner disk winds.  In particular,
\citet{edwardsea03, edwardsea06} observed the He I 10830 \AA\ line in
39 CTTSs and saw several cases with a broad, deep, blue-shifted
absorption, indicating outflow velocities of typically a few hundred,
and up to $\sim 400$ km s$^{-1}$.  They concluded that this feature is
best interpreted as arising in an optically thick stellar wind
\citep[see also][]{dupreeea05}.  They also suggested the winds may be
accretion-powered, since the wind signatures are most prevalent in the
stars with highest accretion rates and absent in non-accreting
systems.  Subsequent modeling of the He I 10830 \AA\ line by
\citet{kwan3ea07} indicates that approximately half of these CTTSs
show evidence for a powerful stellar wind.  Furthermore,
\citet{kurosawa3ea06} modeled the H$\alpha$ emission line in these
systems and suggested that a stellar wind component could most
naturally explain the profiles observed in $\sim 7$\% of the stars in
a sample compiled by \citet{reipurth3ea96}.


There already exists some theoretical work on stellar winds,
specifically from pre-main-sequence stars, with a focus on the wind
driving mechanism \citep{decampli81, hartmann3ea82, hartmannea90} or
the collimation of the winds \citep{fendt3ea95, fendtcamenzind96}.
These do not discuss the expected angular momentum outflow rates,
however.  The works that do calculate stellar wind torques for
pre-main-sequence stars \citep[][Paper I]{hartmannmacgregor82,
  mestel84, hartmannstauffer89, toutpringle92, paatzcamenzind96} are
either based on a 1-dimensional formulation and/or have made a priori
simplifying assumptions regarding the stellar magnetic field
structure, wind flow speed, and latitudinal dependence of the wind.
Calculating the stellar wind torque reliably is a complex,
multi-dimensional problem, and more work is needed to develop the
stellar wind theory further.

The primary goal of this paper therefore, is to take the next major
step in developing the accretion-powered stellar wind picture by
rigourously computing the steady-state solutions of winds from
spinning magnetized stars.  We carry out a parameter study to provide
a range of possible solutions that are expected to charaterize
accretion-powered stellar winds.  Where possible, we compare our
results to analytic magnetohydrodynamic (MHD) stellar wind theory.  In
a companion paper, we will use these solutions to compare the stellar
wind torques and wind driving power with the torque and energy
deposition expected to arise from the interaction of the star with its
accretion disk.

In the following section (\S \ref{sec_theory}), we give a brief
introduction to basic stellar wind theory.  This provides the
motivation for using a numerical approach and sets the stage for
comparing our numerical results with the analytic theory.  Section
\ref{sub_driving} contains a discussion of our adopted wind driving
mechanism.  We describe our numerical method for obtaining solutions
in section \ref{sec_method}, and present the results in section
\ref{sec_solutions}.  Section \ref{sub_rasim} contains a semi-analytic
formulation for the torque and a comparison to previous theory.


\section{Magnetized Stellar Winds: Needed Background}

     \subsection{Magnetic Stellar Wind Theory} \label{sec_theory}

Standard MHD wind theory (i.e., magnetic rotator theory), following
\citet{weberdavis67}, characterizes a steady-state flow of plasma
along a magnetic field line that is anchored to a rotating object,
which we will hereafter take to be a star.  One of the key results is
that the angular momentum outflow rate per unit mass loss is given
very simply as \citep[see, e.g.,][]{weberdavis67, mestel68, michel69}
\begin{eqnarray}
\label{eqn_l}
l = \Omega_* r_{\rm A}^2,
\end{eqnarray}
where $\Omega_*$ is the angular rotation rate of the star, and
$r_{\rm A}$ is the cylindrical radius at which the outflow speed
equals the local magnetic Alfv\'en speed,
\begin{eqnarray}
\label{eqn_va}
v_{\rm A} \equiv {{B_{\rm p}} \over {\sqrt{4 \pi \rho}}},
\end{eqnarray}
where $\rho$ is the local mass density and $B_{\rm p}$ is the strength
of the poloidal magnetic field, $B_{\rm p} = (B_r^2 + B_z^2)^{1/2}$,
in cylindrical ($r$, $\phi$, $z$) coordinates.  Equation (\ref{eqn_l})
indicates that the quantity of angular momentum carried in the wind is
{\it as if} the wind material is corotating out to $r_{\rm A}$ and
conserves its angular momentum thereafter.  Thus, $r_{\rm A}$ is often
referred to the magnetic ``lever arm.''  In reality, the azimuthal
velocity of the wind, $v_\phi$, is a smooth (i.e., differentiable)
function of radius, and the difference between $v_\phi r$ and $l$ at
all radii equals the torque transmitted by azimuthally twisted
magnetic field lines.

By integrating the mass flux times $l$ over any surface enclosing the
star, one obtains an expression for the total angular momentum outflow
rate and, by Newton's third law, the torque on the star:
\begin{eqnarray}
\label{eqn_tw}
\tau_{\rm w} = - \dot M_{\rm w} \Omega_* \left< r_{\rm A}^2 \right>,
\end{eqnarray}
where $\dot M_{\rm w}$ is the integrated wind mass loss rate.  Since
the value of $r_{\rm A}$ will generically not be the same along each
field line, equation (\ref{eqn_tw}) defines the quantity $\left<
r_{\rm A}^2 \right>$, which is the mass-loss-weighted average of
$r_{\rm A}^2$ \citep[suggested by][]{washimishibata93}. Hereafter, we
will simply refer to this average as $r_{\rm A} \equiv \left< r_{\rm
  A}^2 \right>^{1/2}$.

The difficulty now lies in calculating $r_{\rm A}$.  The lever arm
length clearly depends on the stellar surface field strength ($B_*$),
stellar radius ($R_*$), and $\dot M_{\rm w}$ because these directly
affect Alfv\'en condition. But it also depends on the flow speed and
field structure, which are not possible to determine a priori in the
wind.  The flow speed is influenced by the thermal energy in
the wind as well as rotation.  In addition, there exist two different
regimes \citep{belchermacgregor76}: the fast magnetic rotator regime,
where the flow speed is mostly determined by magnetorotational
effects; and the slow magnetic rotator, where the flow speed is solely
determined by the wind driving.  The field structure in the wind, even
though the geometry may be known at the stellar surface, is determined
by the self-consistent interaction between the wind and rotating
magnetic field and thus is a function of all parameters.  Therefore
one can only calculate $r_{\rm A}$ by making a priori assumptions
about the field structure and/or flow speed \citep{weberdavis67,
mestel68, okamoto74, mestel84, mestelspruit87, kawaler88} or by using
iterative techniques \citep[or numerical
simulations;][]{pneumankopp71, sakurai85, washimishibata93,
keppensgoedbloed00, mattbalick04}.

All of these methods are complementary.  The analytical work, in which
the field structure is guessed, produces a predictive formulation of
the stellar wind torque \citep[e.g.,][]{kawaler88}.  However, usually
the formulation of the field structure introduces more parameters
(such as a power law index for the magnetic field), so that almost any
result can be obtained by adjusting these.  Furthermore, the field
structure in the analytic models has no explicit dependence on (e.g.)\
$\Omega_*$, which is exhibited in numerical simulations
\citep[e.g.,][]{mattbalick04}.  The numerical simulation technique has
the advantage of calculating the field structure and flow speed
self-consistently.  However, a single simulation does not predict the
dependence of $r_{\rm A}$ on parameters, and to date, not enough
parameter space has been explored.  Thus, to date, there exists no
formulation for the stellar wind torque that convincingly applies over
a wide range of conditions (e.g., over a range of $B_*$, $\dot M_{\rm
w}$, and $\Omega_*$).

In this paper we will use 2-dimensional (axisymmetric) MHD simulations to
calculate the torque and corresponding value of $r_{\rm A}$.  This
will allow us to check the estimate for $r_{\rm A}$ of Paper I (and
previous works).  In addition, we will carry out a parameter study to
determine the dependence of the stellar wind torque on parameters,
over a range of conditions appropriate for TTSs, and compare with the
predictions of analytic theory.

     \subsection{Wind Driving Mechanism} \label{sub_driving}

It is not known what drives winds from TTSs.  These stars have active
coronae \citep{feigelsonmontmerle99, stassunea04, favataea05}, and it
thus seems a reasonable assumption that they also drive solar-like
coronal winds in which thermal pressure plays a significant role in
the wind acceleration.  Based on a calculation from
\citet{bisnovatyikoganlamzin77}, \citet{decampli81} concluded that, in
order for the wind emission to be consistent with the X-ray
observations, the mass loss rate of a T Tauri star coronal wind must
be less than $\sim 10^{-9} M_\odot$ yr$^{-1}$.  Furthermore,
\citet{dupreeea05} found evidence for a stellar wind with a coronal
temperature in the CTTS TW Hya \citep[though this conclusion has been
challenged by][]{johnskrullherczeg07}.

The assumption of thermal pressure driving is a simplification, even
for the solar wind.  It is known that a major factor in driving the
solar wind is Alfv\'en wave momentum and energy deposition.  Two
important recent studies have done self-consistent analyses of the
combined problem of both solar wind heating and acceleration
\citep{suzukiinutsuka06, cranmer3ea07}.  The first paper shows that
low frequency, transverse motions of open field lines at the
photosphere leads to transonic solar winds for superradial expansion
of the wind cross-section. If the amplitude of these transverse
photospheric motions exceeds 0.7 km s$^{-1}$, fast winds are produced
and the dissipation of wave energy heats the atmosphere to a million
degrees.  The results are sensitive to the amplitude of the velocity
perturbations, and the simulations show that the solar wind vitually
disappears for amplitudes $\le 0.3$ km s$^{-1}$.  These numerical
simulations also show that Alfv\'en wave pressure dominates the gas
pressure in the solar acceleration region ($1.5 R_{\odot} \le R \le 10
R_{\odot}$).  The second paper shows similar results.  This work shows
that there are three key parameters that control wind heating and
acceleration: the flux of acoustic power injected at the photosphere,
the Alfv\'en wave amplitude there, and the Alfv\'en wave correlation
length (characterizing wave damping through turbulence) at the
photosphere.

Our primary goal here is to evaluate the angular momentum transported
away from the star by the stellar wind.  Thus, in this work, we do not
discuss the thermodynamic properties of the wind and instead focus on
the angular momentum transport.  Fortunately, this torque does not
much depend on what drives the wind.  Rather, the torque depends
primarily on the stellar magnetic field, rotation rate, radius, $\dot
M_{\rm w}$, and the wind velocity.  As long as ``something''
accelerates the wind to speeds similar to what we see in our
simulations, the torque we calculate will be approximately correct.

We expect that the Alfv\'en waves in accreting TTS winds will have a
significant, if not dominant contribution to both the acceleration and
heating of their winds.  These waves will be launched along the open
field lines that originate from the TTS photosphere at latitudes
comparable to those that harbour field lines carrying the accretion
flow onto the star.  The irregular accretion flow should generate very
large (i.e., much larger than acoustic motions in the solar
photosphere) acoustical transverse motions in the TTS photosphere as
it impinges upon the star. These large amplitude perturbations,
generated by the accretion flow itself, may be the ultimate driver for
the Alfv\'en wave flux that drives our proposed accretion-powered
stellar wind.

Note that
the driving force can be parameterized as being proportional to
$-\mbox{\boldmath $\nabla$} \xi$ \citep[where $\xi$ is the wave energy
density;][]{decampli81}.  This has the same functional form as the
thermal pressure force ($- \mbox{\boldmath $\nabla$} P$) used in our
simulations.  Several authors \citep[e.g.,][]{hartmannmacgregor80,
decampli81, holzer3ea83, suzuki07} computed velocity profiles for cool
($\sim 10^4$ K) Alfv\'en wave-driven winds.  These works exhibit wind
velocity profiles that are similar to what is expected from thermal
pressure driving of hotter winds.  Therefore, we can think of thermal
pressure driving as a proxy for some other driving mechanism.  Also,
it will be important to have these solutions to compare with future
work that includes different driving mechanisms.

In this paper, we restrict ourselves to mass loss rates of $\dot
M_{\rm w} < 2 \times 10^{-9} M_\odot$ yr$^{-1}$.  As justified above,
we adopt a Parker-like \citep{parker58} coronal wind driving
mechanism, modified by magnetic fields, stellar rotation, and an
enhanced mass loss rate (relative to the sun).  As the nature (e.g.,
temperature) of TTS stellar winds is not well-known, our detailed
solutions of coronal winds will enable us to look at the expected
radiative properties, a posteriori, allowing for further constraints
on real systems.  We will show in a forthcoming paper \citep[and
  see][]{mattpudritz07iau} that the expected emission from the
simulated winds presented here rules out thermal pressure driving at a
substantially lower mass loss rate than the limit of
\citet{decampli81}.

\section{Numerical Simulation Method} \label{sec_method}

We calculate solutions of steady-state winds from isolated stars (no
accretion disk), using the finite-difference MHD code of
\citet{mattbalick04}, and the reader will find further details
there\footnote{\citet{mattbalick04} ran cases with isotropic
hydrodynamic variables at the base of the wind and also cases with
enhanced polar winds.  Here we only consider the isotropic case.}
(and references therein).  Assuming axisymmetry and using a
cylindrical ($r$, $\phi$, $z$) coordinate system, the code employs a
two-step Lax-Wendroff scheme \citep{richtmyermorton67} to solve the
following time-dependent, ideal MHD equations:
\begin{eqnarray}
{{\partial\rho}\over{\partial t}} &=& 
	-\mbox{\boldmath $\nabla$}\cdot(\rho \mbox{\boldmath $v$}), \\
{{\partial(\rho \mbox{\boldmath $v$})}\over\partial t} &=& 
	-\rho(\mbox{\boldmath $v$} \cdot \mbox{\boldmath $\nabla$})
	\mbox{\boldmath $v$}  - \mbox{\boldmath $v$} 
	[\mbox{\boldmath $\nabla$} \cdot (\rho \mbox{\boldmath $v$})]
	\nonumber \\
	& & - \mbox{\boldmath $\nabla$}P
	-{{GM_*\rho}\over {(\mbox{\boldmath $r$}^2+\mbox{\boldmath $z$}^2)}}
	\hat{\mbox{\boldmath $R$}}
	+{{1}\over {\rm c}}(\mbox{\boldmath $J$}\times\mbox{\boldmath $B$}), 
	\label{eq_mom} \\
{{\partial e }\over\partial t} &=& 
	-\mbox{\boldmath $\nabla$} \cdot [\mbox{\boldmath $v$}(e + P)]
	-\left[{{GM_*\rho}\over 
	{(\mbox{\boldmath $r$}^2+\mbox{\boldmath $z$}^2)}}
	\hat{\mbox{\boldmath $R$}}\right]\cdot\mbox{\boldmath $v$}
	+\mbox{\boldmath $J$} \cdot \mbox{\boldmath $E$}, \label{eq_eng}\\
{{\partial\mbox{\boldmath $B$}}\over\partial t} &=& 
	-{\rm c}(\mbox{\boldmath $\nabla$}\times\mbox{\boldmath $E$}),
\end{eqnarray}
and uses
\begin{eqnarray}
\mbox{\boldmath $E$} &=& 
  -{{1}\over{\rm c}}(\mbox{\boldmath $v$}\times\mbox{\boldmath $B$}), \\
\mbox{\boldmath $J$} &=& 
  {{\rm c}\over {4\pi}}(\mbox{\boldmath $\nabla$}\times\mbox{\boldmath $B$}),\\
e 	&=& 
  {{1}\over {2}}\rho v^2 + {{P}\over {\gamma - 1}},
\end{eqnarray}
where $\rho$ is the density, \mbox{\boldmath $v$} the velocity, $P$
the gas pressure, $G$ Newton's gravitational constant, $M_*$ the
stellar mass, $R$ the spherical distance from the center of the star
($R^2 = r^2 + z^2$), $e$ the internal energy density, \mbox{\boldmath
$B$} the magnetic field, \mbox{\boldmath $J$} the volume current,
\mbox{\boldmath $E$} the electric field, c the speed of light, and
$\gamma$ the ratio of specific heats.

To obtain steady-state wind solutions, we follow the method of
\citet{mattbalick04}, which is also similar to that employed by
\citet{washimishibata93} and \citet{keppensgoedbloed99}.  It involves
initializing the computational grid with a spherically symmetric,
isothermal Parker wind solution \citep{parker58}, plus force-free
dipole (and sometimes quadrupole) magnetic field.  When the simulation
begins, the wind solution changes from the initial state due to the
presence of the magnetic field, the rotation of the star, and the
polytropic equation of state ($P \propto \rho^\gamma$).  The
simulations run until the system relaxes into a steady-state (within a
small tolerance) MHD wind solution.  The code uses nested
computational grids so that the wind can be easily followed to large
distances (several tens to hundreds of $R_*$).

This method results in a steady-state solution for the wind that is
determined solely by the boundary conditions held fixed at the base of
the stellar corona (the ``stellar surface").  In order to capture the
appropriate physics within the framework of a finite difference
scheme, we employ a four-layer boundary for the star, on which the
various physical quantities are set as follows.  We consider the
spherical location $R = 30$, in units of the grid spacing, to be the
surface of the star. For all gridpoints such that $R \le 34.5$, the
poloidal velocity is forced to be parallel with the poloidal magnetic
field ($v_{\rm p} \parallel B_{\rm p}$, where the poloidal component
is defined as the vector component in the $r$-$z$ plane).  Where $R
\le 33.5$, $\rho$ and $P$ are held constant (in time) at their initial
values.  For $R \le 32.5$, $v_{\rm p}$ is held at zero, while $v_\phi$
is held at corotation with the star.  For $R \le 31.5$, $B_{\rm p}$
field is held at its initial, dipolar value, while $B_\phi$ is set so
that there is no poloidal electric current at that layer (which gives
it a dependence on the conditions in the next outer layer, $31.5 < R
\le 32.5$).

These boundary conditions properly capture the behavior of a wind
accelerated from the surface of a rotating magnetized star, as
follows.  There is a layer on the stellar boundary ($R > 32.5$)
outside of which the velocity not fixed, but is allowed to vary in
time.  In this way, the wind speed and direction is not specified, but
is determined by the code in response to all of the forces.  By
holding $P$ fixed at its initial value for all $R \le 33.5$, we
constrain the pressure gradient force (thermal driving) at the base of
the wind to be constant in time.  Also, holding the density fixed at
$R \le 33.5$ allows the region from where the wind flows to be
instantly replenished with plasma.  Thus, the base of the wind
maintains a constant temperature and density, regardless of how fast
or slow the wind flows away from that region.  The existence of a
layer in which $v_{\rm p} = 0$ and $B_{\rm p}$ can evolve (namely, at
$31.5 < R \le 32.5$) allows $B_{\rm p}$ (and $v_{\rm p}$) to reach a
value that is self-consistently determined by the balance of magnetic
and inertial forces.  We set the poloidal velocity parallel to the
poloidal magnetic field for the next two outer layers, to ensure a
smooth transition from the region of pure dipole field and zero
velocity to a that with a perturbed field and outflow.  Setting
$B_\phi$ so that the poloidal electric current is zero inside some
radius ensures that the field behaves as if anchored in a rotating
conductor (the surface of the star).  Also, this ensures that $B_\phi$
evolves appropriately outside the anchored layer according to the
interaction with the wind plasma.

The key physical parameters can be represented by the characteristic
speeds of the input physics, namely the sound speed at the base of the
corona, $c_{\rm s}$, the escape speed from the surface of the star,
$v_{\rm esc}$, the rotation speed of the star, and the Alfv\'en speed
at the base of the wind.  We specify the ratio of $c_{\rm s} / v_{\rm
  esc}$ as our parameter, rather than the sound speed alone.  This
seems the most reasonable, since the temperature of a thermally driven
wind is regulated somewhat by the interplay between the thermal energy
input and the expansion of the corona (the wind) against gravity.  To
first order, a hotter wind expands more rapidly against gravity
allowing less time for the gas to heat, and a cooler wind expands more
slowly, allowing more time to heat.  Once the value of the stellar
mass and radius is specified, the ratio of $c_{\rm s} / v_{\rm esc}$
determines the temperature held fixed on the stellar boundary, as
described above.  The wind plasma is characterized by a polytropic
equation of state, and so $\gamma$ is also a parameter.  We
parameterize the stellar rotation rate as the fraction of breakup
speed, 
\begin{eqnarray}
f \equiv \Omega_* R_*^{3/2} (G M_*)^{-1/2}.
\end{eqnarray}

The Alfv\'en speed is determined by the magnetic field strength and
coronal density.  Rather than taking the Alfv\'en speed as a key
parameter, we specify the field strength at the equator of the star
($B_*$) as our parameter, in order to connect the simulations as much
as possible to observationally constrained quantities.  For the same
reason, we specify $\dot M_{\rm w}$ as a parameter, rather than the
coronal density.  In the simulations, we must specify the base
density, $\rho_*$, to be held fixed on the stellar boundary, and the
value of $\dot M_{\rm w}$ in the steady-state wind is not solely
determined by $\rho_*$.  For example, the rotation of the star can
enhance $\dot M_{\rm w}$ via magneto-centrifugal flinging, and a
strong magnetic field can decrease $\dot M_{\rm w}$ by inhibiting flow
from a region near the equator that remains magnetically closed (the
``dead zone'').  In other words, $\dot M_{\rm w}$ is not an a priori
tunable parameter; rather, it is a result of the simulations.
Therefore, to treat $\dot M_{\rm w}$ as our tunable parameter, we
adopt an iterative approach.  This entails first running a given
simulation with a guess for $\rho_*$, checking the resulting value of
$\dot M_{\rm w}$, and then adjusting $\rho_*$ and rerunning the
simulation.  We iterate until the desired value of $\dot M_{\rm w}$ is
achieved (within a tolerance of 2\%).  This typically required 2 to 4
iterations, so the ability to treat $\dot M_{\rm w}$ as a chosen
parameter comes at a substantial cost.


\section{Stellar Wind Solutions} \label{sec_solutions}

     \subsection{The Fiducial Case} \label{sub_fiducial}

\begin{deluxetable}{ll}
\tablewidth{0pt}
\tablecaption{Fiducial Stellar Wind Parameters \label{tab_parms}}
\tablehead{
\colhead{Parameter} &
\colhead{Value}
}

\startdata

$M_*$                     & 0.5 $M_\odot$ \\
$R_*$                     & 2.0 $R_\odot$ \\
$B_*$ (dipole)            & 200 G         \\
$f$                       & 0.1           \\
$\dot M_{\rm w}$\tablenotemark{a} & $1.9 \times 10^{-9} M_\odot$ yr$^{-1}$ \\
$c_{\rm s} / v_{\rm esc}$ & 0.222          \\
$\gamma$                  & 1.05           \\

\enddata

\tablenotetext{a}{In order to treat $\dot M_{\rm w}$ as a parameter in 
the simulations, our method is to adjust the mass density at the base 
of the wind until the desired $\dot M_{\rm w}$ is achieved in the 
steady-state.}

\end{deluxetable}

We start by presenting the results of our stellar wind simulation for
parameters with values that represent a ``typical'' T Tauri star
and follow the fiducial values of Paper I and
\citet{mattpudritz05}.  Table \ref{tab_parms} lists the fiducial
parameters.  We consider a low mass pre-main-sequence star, with a
surface escape speed of $v_{\rm esc} \approx 309$ km s$^{-1}$.  A
dipole magnetic field strength of 200 Gauss is consistent with $3
\sigma$ upper limits \citep{johnskrullea99, smirnovea04, smirnov3ea03}
or marginal detection \citep{smirnovea03, yang3ea07} of the
longitudinal magnetic field measured for CTTSs.  We seek primarily to
understand the slow rotators, for which a rotation rate of 10\% of
breakup is appropriate.  In Paper I, we estimated that an
accretion-powered stellar wind for a T Tau star might have
$\dot M_{\rm w} \approx 1.9 \times 10^{-9} M_\odot$ yr$^{-1}$, so we
use this as our fiducial value.

In a thermally-driven wind, the coronal sound speed should be
comparable to the escape speed, and we use $c_{\rm s} / v_{\rm esc} =
0.222$ as our fiducial value.  This value gives wind speeds that are
appropriate in the solar case.  The choice of polytropic index
$\gamma$ is also important.  At large distances ($\sim$AU) from the
sun, the solar wind plasma is well characterized by an effective
$\gamma$ between approximately 1.5 and 5/3 \citep{feldmanea98,
  krasnopolsky00}.  However, in the region where the wind is
accelerated (within a few solar radii), thermal conduction and other
heating and cooling effects play a role \citep[e.g.,][]{cranmer3ea07},
resulting in an effective $\gamma$ closer to unity (isothermal).  Our
fiducial value of $\gamma = 1.05$ was used by \citet{washimishibata93}
and \citet{mattbalick04} for solar-like winds.  This nearly isothermal
value of $\gamma$ approximates the thermodynamics of a gas with a true
value of $\gamma = 5/3$ that is heated as it expands.


\begin{figure}
\epsscale{1.15}
\plotone{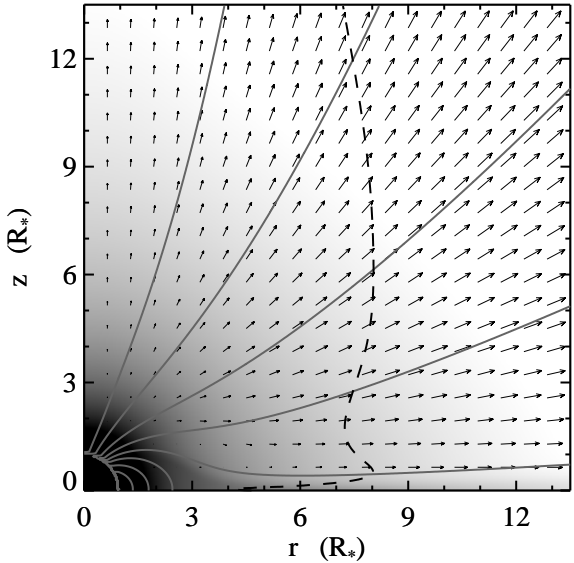}

\caption{Fiducial case: greyscale of log density, velocity vectors,
  and magnetic field lines illustrate the structure of the
  steady-state wind solution (see table \ref{tab_parms}).  The dashed
  line represents the Alfv\'en surface, where the wind speed equals
  the local Alfv\'en speed.  The rotation axis is vertical, and the
  longest vector corresponds to 160 km s$^{-1}$.  Black corresponds to
  a density above $5.3 \times 10^{-13}$ g cm$^{-3}$ and white to a
  density below $2.6 \times 10^{-16}$ g cm$^{-3}$.
\label{fig_fiducial}}

\end{figure}

Figure \ref{fig_fiducial} shows the result of our fiducial case
simulation, which illustrates the steady-state wind solution.  The
non-spherical shape of the Alfv\'en surface (which eventually crosses
the rotation axis at larger radii than shown) is mainly due to
magnetorotational effects in the wind \citep[see][]{washimishibata93,
mattbalick04}.  This demonstrates that the fiducial T Tauri star wind
exists in the thermo-centrifugal regime where thermal and
magnetocentrifugal effects are of similar importance for accelerating
the wind \citep{sakurai85, washimishibata93}.  These winds are
self-collimated, while still exhibiting substantial flow at all
latitudes.

From the simulation, we calculate $\dot M_{\rm w}$ and the total
angular momentum outflow rate, $\tau_{\rm w}$, as described by
\citet{mattbalick04}.  Then, using equation (\ref{eqn_tw}), we
calculate the effective lever arm length, $r_{\rm A} \equiv \left<
r_{\rm A}^2 \right>^{1/2}$.  These results are listed in the first row
of table \ref{tab_simresults}, where we also list the coronal base
density $\rho_*$ that we iteratively chose to give the desired value
of $\dot M_{\rm w}$.

\begin{deluxetable}{lcccc} 
\tablewidth{0pt}
\tablecaption{Stellar Wind Torques and Lever Arm Lengths \label{tab_simresults}}
\tablehead{
\colhead{Case} &
\colhead{$\rho_*$} &
\colhead{$\dot M_{\rm w}$} &
\colhead{$\tau_{\rm w}$} &
\colhead{$\left< r_{\rm A}^2 \right>^{1/2}$}  \\
\colhead{} &
\colhead{($10^{-11} {{\rm g} \over {\rm cm}^{3}}$)} &
\colhead{($10^{-9} {M_\odot \over {\rm yr}}$)} &
\colhead{($10^{36}$ erg)} &
\colhead{$(R_*)$} 
}

\startdata

fiducial              & 3.67  & 1.89 & 1.77   & 6.97  \\
$f$ = 0.004           & 7.62  & 1.86 & 0.0972 & 8.33  \\
$f$ = 0.2             & 1.36  & 1.87 & 2.82   & 6.26  \\
$f$ = 0.05            & 6.01  & 1.88 & 1.06   & 7.65  \\
$B_*$ = 400 G         & 3.67  & 1.86 & 3.27   & 9.55  \\
$B_*$ = 2 kG          & 3.67  & 1.92 & 13.8   & 19.3  \\
1 kG quad.            & 2.92  & 1.87 & 1.37   & 6.17  \\
2 kG quad.            & 4.38  & 1.93 & 2.11   & 7.53  \\
low $\dot M_{\rm w}$  & 0.377 & 0.187 & 0.500 & 11.8  \\
very low $\dot M_{\rm w}$ & 0.0755 & 0.0378 & 0.204 & 16.7 \\
$R_*$ = 1.5 $R_\odot$ & 5.71  & 1.86 & 1.10   & 5.96  \\
$R_*$ = 3 $R_\odot$   & 1.99  & 1.89 & 3.43   & 8.75  \\
$M_*$ = 0.25 $M_\odot$& 5.06  & 1.91 & 1.47   & 7.52  \\
$M_*$ = 1 $M_\odot$   & 2.59  & 1.88 & 2.11   & 6.42  \\
$c_{\rm s}/v_{\rm esc}$ = 0.245 & 0.773 & 1.87 & 1.59 & 6.64 \\
$c_{\rm s}/v_{\rm esc}$ = 0.192 & 55.4  & 1.89 & 1.91 & 7.23 \\
$\gamma$ = 1.10       & 11.1  & 1.87 & 2.19   & 7.79 \\

\enddata

\end{deluxetable}

The wind base density of $\sim 10^{-11}$ gm cm$^{-3}$ is 5 orders of
magnitude larger than required for simple solar wind models
\citep[e.g.,][]{washimishibata93}.  This is expected, since the
fiducial $\dot M_{\rm w}$ is 5 orders of magnitude higher than the
solar value, and the wind speeds are comparable.

The fiducial stellar wind torque of $\approx 1.8 \times 10^{36}$ ergs
is capable of balancing the spin up torque from accretion at a rate of
$4.4 \times 10^{-9} M_\odot$ yr$^{-1}$.  The basic conclusion here is
that the stellar wind torque for the fiducial case is of the right
magnitude to be important for spinning down the star, as required by
the accretion-powered stellar wind scenario.  We chose our fiducial
parameters to compare with the estimate of Paper I that\footnote{Paper
I actually quotes a value of $r_{\rm A} / R_*$ = 15, but our
definition differs slightly here (compare eq.\ \ref{eqn_tw} here with
eq.\ 2 of Paper I), so the lever arm length corresponds to 12.2 $R_*$
here.} $r_{\rm A} / R_* \approx 12.2$.  We can see that their
estimate, based on scaling of 1D wind theory from solar values, was a
75\% overestimate of $r_{\rm A}$.  We will identify the reasons for
this in section \ref{sec_parmspin}.


     \subsection{Parameter Study}

To establish the dependence of $r_{\rm A}$ on parameters and to
calculate wind solutions that are applicable to a wide range of
conditions that are observed or often assumed for T Tauri stars, we
carried out a limited parameter study with our simulations.  The
results are listed in table \ref{tab_simresults}, and we briefly
discuss each case below.  The first column in the table lists the
value of the parameter that is changed relative to the fiducial case.
For each case, all other parameters are identical to the fiducial
case.  Note that since we consider $\dot M_{\rm w}$ as a key
parameter, the value of $\rho_*$ varies from case to case.

          \subsubsection{Spin Rate} \label{sec_parmspin}

\begin{figure}
\epsscale{1.15}
\plotone{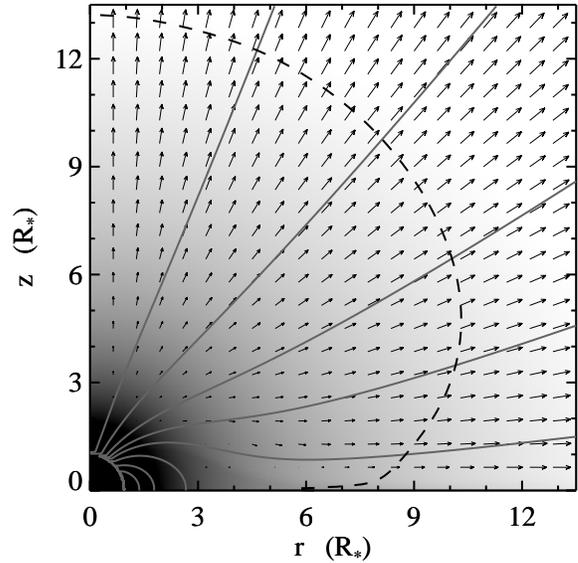}

\caption{Same as figure \ref{fig_fiducial}, but for the $f = 0.004$
(similar to solar) case.  The maximum velocity vector corresponds to 95 km
s$^{-1}$. \label{fig_f004}}

\end{figure}

As with the fiducial case, all but one simulation in our parameter
study lie in a regime that is near the boundary between slow and fast
magnetic rotators.  The one exception is a case with a fractional
rotation rate equal to the solar value of $f = 0.004$, which
represents a slow magnetic rotator.  Figure \ref{fig_f004} illustrates
the structure of the steady-state wind solution for this case in the
same format as the previous figure.  A comparison between the two
figures reveals that rotation indeed influences the detailed structure
of the velocity field and the magnetic field in the wind, which
manifests itself as a difference in the shape of the Alfv\'en surface.
All else being equal, the effect of faster rotation is to reduce the
effective lever arm length, as evident in table \ref{tab_simresults}
\citep[see also][]{sakurai85, washimishibata93}. Although the
qualitative effect of rotation on the shape of the Alfv\'en surface
was anticipated in analytic theory
\citep[e.g.,][]{belchermacgregor76}, this effect is not properly
included in any existing analytic formulation for calculating the
torque. This is a primary reason that numerical simulations are
required to convincingly calculate the self-consistent wind solution,
especially when considering winds that exist near the boundary between
slow and fast magnetic rotators.  The effect of rotation on $r_{\rm
  A}$ was not considered in the estimate of Paper~I and accounts for
approximately 30\% of their overestimate of $r_{\rm A}$.

As described in \citet{mattbalick04}, a given simulation can be scaled
to other systems with the same characteristic velocity ratios, so that
the resulting value of $r_{\rm A} / R_*$ is valid for a family of
solutions.  The simulation with $f = 0.004$ scales to a solution very
similar to the solar wind with $R_* = 1 R_\odot$, $M_* = 1 M_\odot$,
$\dot M_{\rm w} = 1.3 \times 10^{-14} M_\odot$ yr$^{-1}$, $B_* = 1.5$
G, and all speeds are increased by a factor of 2.  Thus, for these
parameters, this simulation predicts a lever arm length of 8.33
$R_\odot$ for the case of the solar wind, and $\tau_{\rm w} = 6.8
\times 10^{29}$ erg.  This torque is consistent with the numerical
results of \citet{washimisakurai93}, but a factor of a few times
smaller than observationally determined values \citep{li99}.  To
obtain the observed solar torque, corresponding to $r_{\rm A} = 12.2
R_\odot$, the simulation would require (e.g.) a substantially stronger
magnetic field than 1 G.  This was also suggested by \citet{li99}, and
our simulations corroborate that suggestion.  As the estimate in Paper
I assumed the canonical value of $B_* \approx 1$ G for the sun, this
accounts for most of the discrepancy between our simulation results
and the Paper I estimate of $r_{\rm A}$.

To capture a range of spins appropriate for the T Tauri star
``slow rotators,'' we also ran cases with spin rates of twice and half
of the fiducial spin rate.  The results of these simulations are
listed in the 3rd and 4th row of table \ref{tab_simresults}.

          \subsubsection{Dipole Field Strength}

\begin{figure}
\epsscale{1.15} 
\plotone{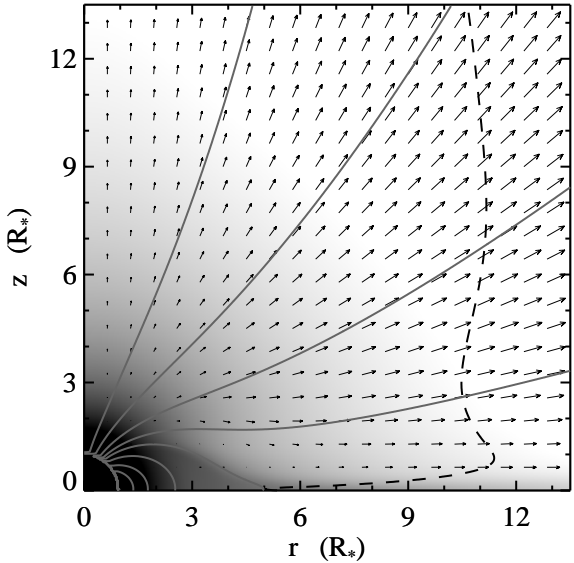}

\caption{Same as figure \ref{fig_fiducial}, but for the $B_* = 400$
Gauss dipole case.  The maximum velocity vector corresponds to 190 km
s$^{-1}$. \label{fig_b400g}}

\end{figure}

Measurements for the mean $|B_*|$ exist for a number of T Tauri stars
\citep[e.g.,][]{johnskrull07}.  These results show a remarkably
consistent field strength for all stars of around 2 kG.  Measurements
of the longitudinal field (which limits the global, dipole component)
exist only for a handful of accreting stars \citep{bouvierea07,
johnskrull07iau}.  These measurements suggest the dipole component is
no greater than 200 G (though larger values would be allowed for
special viewing geometries).  Given the small number of measurements,
it is still relevant to consider stronger dipole field strengths.
Thus we have run cases with $B_*$ = 400 and 2 kG.

Figure \ref{fig_b400g} illustrates the wind solution for the case with
$B_* = 400$ G, and the results of both cases are listed in table
\ref{tab_simresults}.  It is clear that the strength of the field has
a strong influence on the stellar wind torque.

	  \subsubsection{Surface Field Geometry} \label{sub_geometry}

\begin{figure}
\epsscale{1.15}
\plotone{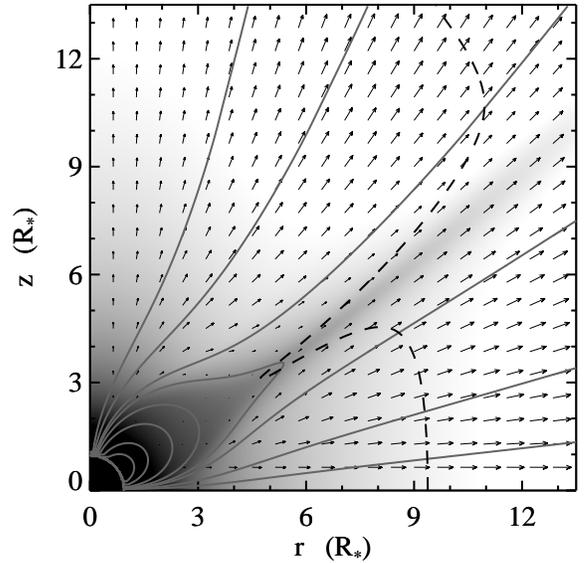}

\caption{Same as figure \ref{fig_fiducial}, but for the 2 kG
quadrupole case.  The maximum velocity vector corresponds to 170 km
s$^{-1}$. \label{fig_quad}}

\end{figure}

The fact that T Tauri stars have a mean field of $|B_*| \sim 2$ kG
with a much weaker dipole component, indicates that the stellar
surface field is dominated by higher order multipole fields.
Therefore, it may be important for future work to include much more
structured fields than we consider here.  To begin to quantify the
effects of higher order fields, we ran two cases that were initialized
with a quadrupolar field of the form
\begin{eqnarray}
\label{eqn_quad}
B_R &=& B_* (3 \cos^2 \theta - 1) 
        \left({R_* \over R}\right)^4  \nonumber \\ 
B_\theta &=& 2 B_* \cos \theta \sin \theta \left({R_* \over R}\right)^4
\end{eqnarray}
in spherical coordinates. 

We ran cases with $B_*$ equal to 1 kG and 2 kG, listed as ``1 kG
quad.'' and ``2 kG quad.'' respectively in table \ref{tab_simresults}.
Figure \ref{fig_quad} illustrates the wind solution for the 2 kG
quadrupole case.  It is clear from the figure that, compared to the
cases with a dipole field, the shape of Alfv\'en surface is quite
different.  Also, for a given value of $B_*$, the effective lever arm
length is much shorter for the quadrupole case.

It is evident from table \ref{tab_simresults} that the stellar wind
torque from a star with a 200 G dipole field is comparable to that
with a 1--2 kG quadrupole field.  Thus, the stellar wind torque is
very sensitive to surface field geometry.  However, since the measured
surface field strengths of $\sim 2$ kG likely include contributions
from even higher multipoles than a quadrupole, it seems that the
dipole component will generally dominate near the Alfv\'en surface.
So the strength of the dipole component should generally be the most
important for determining the torque.

	  \subsubsection{Mass Loss Rate}

We have thus far considered quite massive winds, motivated by recent
suggestions in the literature for accretion powered winds.  However,
the value of $\dot M_{\rm w}$ is very uncertain and is likely to
exhibit a wide range in values from one object to the next.  In
addition, it would be interesting to predict what torques may be
expected for the winds from the non-accreting, weak line TTSs.  We
expect these stars to have solar-like winds that are quite enhanced
relative to their main sequence counterparts, yet probably less
powerful than winds from the accreting stars.

Unfortunately, our method is limited to cases with lever arms that are
not too long, since longer lever arms requires a larger Alv\'en speed
on the stellar surface.  A large Alfv\'en speed increases the time for
the simulation to run and also increases the error in the solution
(e.g., by increasing the effective diffusion rate in our code).  For
this practical reason, we were limited to running only two cases with
lower $\dot M_{\rm w}$ covering a range in $\dot M_{\rm w}$ of a
factor of 50.  These are listed in the 9th and 10th rows of table
\ref{tab_simresults}.

	  \subsubsection{Stellar Radius}

T Tauri stars contract as they age, so stars of a given mass exhibit a
range of radii during this phase.  Thus, it is important to consider
here different combinations of stellar mass and radius.  Table
\ref{tab_simresults} contains results from two cases with $R_* = 1.5
R_\odot$ and $R_* = 3 R_\odot$.  Note that changing $R_*$ changes
$v_{\rm esc}$, so these cases have a different coronal temperature and
$\Omega_*$, in order that $c_{\rm s} / v_{\rm esc}$ and $f$ are
constant.  From the values in the table, it is evident that the
stellar wind torque is very sensitive to $R_*$.  The reason for this
is twofold.  First, since $B_*$ is fixed, a larger stellar radius
corresponds to a larger dipole moment ($\mu \equiv B_* R_*^3$), which
is capable of conveying a larger torque.  Second, a larger stellar
radius decreases the surface gravity, and so the influence of the
magnetic field relative to gravity is increased (i.e., $v_{\rm A} /
v_{\rm esc}$ increases).  Thus, $(r_{\rm A} / R_*)^2$ increases with
$R_*$, and though $\Omega_*$ decreases (to keep $f$ fixed), the
quantity $\Omega_* R_*^2$ increases, so the net torque increases.

	  \subsubsection{Stellar Mass}

Cases with half and twice the fiducial stellar mass are also listed in
table \ref{tab_simresults}.  As with the cases of different $R_*$,
note that changes in $M_*$ change $v_{\rm esc}$, so we have adjusted
the coronal temperature and $\Omega_*$ to keep the parameters listed
in table \ref{tab_parms} fixed.  As with the case of varying $R_*$, a
change in $v_{\rm esc}$ changes the relative importance of the
magnetic field with the gravity.  Thus, $r_{\rm A} / R_*$ is larger
for a smaller $M_*$.  However, since we have fixed $f$, a smaller
$M_*$ means a smaller $\Omega_*$ so that the net stellar wind torque
decreases.

	  \subsubsection{Wind Acceleration} \label{sub_accel}

The increase of the wind speed with distance from the star depends on
the details of the wind acceleration mechanism.  In a Parker wind, the
temperature (parameterized by $c_{\rm s} / v_{\rm esc}$) and the
cooling/heating of the gas as it flows (parameterized by $\gamma$) are
the key physical properties determining the velocity profile in the
wind.  A hotter wind accelerates more rapidly and achieves a higher
speed than a cooler wind.  A wind with a larger $\gamma$ (closer to
5/3) cools more rapidly as it expands, and so the bulk of the
acceleration takes place closer to the star.  Similarly, if the wind
is instead accelerated by something other than thermal pressure, the
velocity profile may be altered.  

In order to quantify the effect of varying the acceleration in the
wind, within the framework of the pressure-driving mechanism used
here, we have run three more simulations.  The results of these are
listed in the last 3 rows of table \ref{tab_simresults}.  In order,
these represent winds that are hotter, colder, or with less heating
(i.e., more adiabatic cooling) than the fiducial case.  The relatively
large effect of $c_{\rm s} / v_{\rm esc}$ and $\gamma$ on the wind
speed near the stellar surface is evident by the very different values
of $\rho_*$ required to keep $\dot M_{\rm w}$ fixed, listed in table
\ref{tab_simresults}.  The wind velocity at the base of the corona,
for fixed $\dot M_{\rm w}$, varies as the inverse of the variation in
$\rho_*$.  So these cases represent large differences in the wind
acceleration rate.  The effect on the torque is relatively small, but
is not entirely negligible.  It will be important for future work to
determine the wind torques for different driving mechanisms.  The
preliminary conclusion to be drawn from this work is that the wind
velocity profile, and therefore wind driving mechanism, does not have
a large effect on the torque.

\section{Semi-Analytic Fit for the Effective Alfv\'en Radius} \label{sub_rasim}

In section \ref{sec_theory}, we pointed out that no reliable
formulation exists for predicting the Alfv\'en radius (and therefore
torque) in a stellar wind from fundamental parameters.  However our
parameter study, even though somewhat limited, can be used to provide
a numerically based approach to this question.  We will use the result
future work, and it also will be of general interest for other stellar
wind studies.

In a 1-dimensional theory, one can assume that the magnetic field strength
approximately follows a single power law of the form $B = B_* (R_* /
R)^n$.  Then the condition that the wind speed equals the Alfv\'en
speed at $r_{\rm A}$ gives \citep[e.g.,][]{kawaler88, toutpringle92}
\begin{eqnarray}
\label{eqn_ra}
 \left({r_{\rm A} \over R_*}\right)^{2n-2} = 
  {{B_*^2 R_*^2} \over {\dot M_{\rm w} v_{r \rm A}}},
\end{eqnarray}
where $v_{r \rm A}$ is the wind speed at the Alfv\'en radius.  There are a 
number of problems.
First, the true magnetic field strength in a wind does
not follow a single power law \citep[e.g.,][]{mestelspruit87}.  Second,
the Alfv\'en surface is neither a sphere nor a cylinder and a spherical
model is quite misleading.
Third, and perhaps most vexing, is the fact that 
$v_{r \rm A}$ has different values at different points along the 
Alfv\'en surface and cannot be determined a priori.  Finally, there is no
explicit dependence of $r_{\rm A}$ on the spin rate or driving
properties of the wind, which we also know to be false.

\begin{figure}
\epsscale{1.15}
\plotone{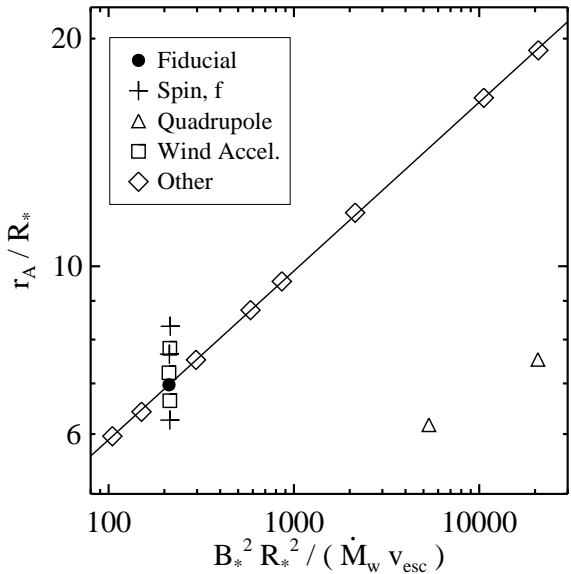}

\caption{Effective lever arm length in the stellar wind versus the
quantity in brackets in equation (\ref{eqn_rasim}).  Shown are the
results of our entire parameter study including the fiducial case
(filled circle); cases with different spin rates (pluses); cases with
a quadrupole field (triangles); cases with different $c_{\rm s} /
v_{\rm esc}$ or $\gamma$ (squares); and all other cases (diamonds),
representing those with different values of $B_*$, $R_*$, $\dot M_{\rm
w}$, or $M_*$.  The line represents the best fit to the fiducial and
``other'' cases, given by equation (\ref{eqn_rasim}) with $K \approx
2.11$ and $m \approx 0.223$
\label{fig_rasim}}

\end{figure}

There is, however, a more general way of scaling the Alfv\'en radius
that is suggested by basic theory.  Another clue is that since the
numerical simulations are carried out in normalized units, they are
scalable to any system with with similar characteristic speeds on the
stellar surface \citep[e.g.,][]{mattbalick04}.  This suggests that we
can replace $v_{r \rm A}$ with the stellar surface escape speed and
calculate the Alfv\'en radius using
\begin{eqnarray}
\label{eqn_rasim}
{r_{\rm A} \over R_*} = K \left({{B_*^2 R_*^2} \over {\dot M_{\rm w}
  v_{\rm esc}}}\right)^m,
\end{eqnarray}
where $K$ and $m$ are dimensionless constants.  The quantity inside
the bracket measures the effective magnetization of the wind, is
similar to that used by \citet[][see their eq.\ 7]{uddoulaowocki02},
and arises naturally in disk wind theory \citep[e.g., equation 2.27
  of][]{pelletierpudritz92}.  It is also a quantity that can be fixed
by observations of stellar properties and wind parameters.

In figure
\ref{fig_rasim}, we plot $r_{\rm A} / R_*$ as a function of the
quantity in brackets in equation (\ref{eqn_rasim}) on a $\log$--$\log$
scale, for all 17 of our simulations.  Since we know equation
(\ref{eqn_rasim}) does not properly include the effects of stellar
rotation or the wind driving mechanism, we calculate the best fit $K$
and $m$ to the fiducial case and only those cases with variations on
$B_*$, $R_*$, $\dot M_{\rm w}$, and $M_*$.  The fit, giving $K \approx
2.11$ and $m \approx 0.223$, is plotted as a line in the figure.

It is remarkable that this fit matches all of the relevant simulations
(filled circle and diamonds in the figure) to an accuracy of less than
one percent.  This is at the level of precision of the numerical
method \citep{mattbalick04}.  Remember that we have taken $c_{\rm
s}/v_{\rm esc}$ and $f$ as our parameters, so that cases with a
different value of $v_{\rm esc}$ (i.e., those with different $R_*$ and
$M_*$) actually also have different wind temperatures (i.e., $c_{\rm
s}$) and stellar angular spin rates ($\Omega_*$) than the fiducial
case.  If we had chosen $\Omega_*$ and $c_{\rm s}$ as our fixed
parameters, there would be a lot more scatter of the diamonds around
the line in figure \ref{fig_rasim}.  Furthermore, our simulations
self-consistently capture the interaction between the stellar wind,
magnetic field, and rotation, without resorting to assumptions about
(e.g.)\ the extent of the dead zone, the dependence of magnetic field
strength with radius, or latitudinal variations in wind quantities.
Therefore, our semi-analytic formulation appears to be an improvement
over existing theory.

Equation (\ref{eqn_rasim}) does have some limitations.  Neither the
previous analytic formulations nor our own semi-analytic approach
properly includes the effects of varying stellar rotation or wind
driving (as evident in figure \ref{fig_rasim}).  As an illustrative
example, in a smaller parameter study with a rotation rate comparable
to the solar rate (not presented here), we found that $K \approx 3.0$
and $m \approx 0.19$.  Also, note that a line connecting the two
points with a quadrupole field suggests $K \approx 1.7$ and $m \approx
0.15$, for these cases.  Thus, the indicies $K$ and $m$ are quite
sensitive to the field geometry and have a smaller (but
non-negligible) sensitivity to changes in the stellar spin rate and
the wind acceleration rate/mechanism.  We leave the precise
determination of the sensitivity of $r_{\rm A}$ to these parameters
for future work.



We can now combine equations (\ref{eqn_tw}) and
(\ref{eqn_rasim}) to get a formula for the stellar wind torque,
\begin{eqnarray}
\label{eqn_twsim}
\tau_{\rm w} = {K^2 \over \sqrt{2}} ~ f ~ v_{\rm esc}^{1-2m} ~
               \dot M_{\rm w}^{1-2m} ~ R_*^{1+4m} ~ B_*^{4m},
\end{eqnarray}
though we know this does not properly contain the dependence
(e.g.)\ on $f$.  This equation is essentially the same as that derived
by \citet{kawaler88}, except for the value of the the dimensionless
constant out front and of the expected value of the exponent parameter
$m$.  The constant is not so crucial, and usually this can be
calibrated to the solar wind torque for a predictive theory (though we
have not done this here).  On the other hand, the value of $m$ is of
far greater importance for predicting the torque for a range of
parameters.

In particular, authors typically have chosen a power law such that the
stellar wind torque is nearly or completely independent of $\dot
M_{\rm w}$ (effectively, $m = 0.5$), which results in $\tau_{\rm w}
\propto B_*^2$ \citep[e.g.,][]{kawaler88, pinsonneaultea89,
  barnessofia96, bouvier3ea97}.  Our basic understanding of the
observed Skumanich-style \citep{skumanich72} spin down of main
sequence stars ($\Omega_* \propto t^{-1/2}$), as well as the expected
dependence of magnetic field strength with rotation rate
\citep{belchermacgregor76}, appears to rely on this or a similar
formulation.  Our fit value of $m \approx 0.223$ gives approximately
$\tau_{\rm w} \propto B_*^{0.9}$ \citep[this was also found
  by][]{washimishibata93}, which is substantially different.

In order to understand the difference between our value of the
exponent and that used by others, it is instructive to consider the
power law formulation of the magnetic field used to derive equation
(\ref{eqn_ra}).  It is generally expected \citep[e.g.,][]{mestel84,
  mestelspruit87, kawaler88} that when the surface magnetic field is
dipolar, as we are considering here, the effective power law index of
the magnetic field in the flow will lie somewhere between the value
for a dipole ($n = 3$) and that for a split monopole ($n=2$).  This
has been the primary justification for the power laws used in the
literature.  Indeed our simulations display the expected behavior of
exhibiting a dipolar geometry near the star and an approximately
monopolar geometry far from the star (e.g., figure
\ref{fig_fiducial}).  However, by comparing equations (\ref{eqn_ra})
and (\ref{eqn_rasim}), we see that our fit value of $m \approx 0.223$
seems to imply a magnetic power law of $n \approx 3.2$.

It is important to realize that the divergence of the magnetic field
in the flow, captured by the power law index $n$, is not the only
important effect, and this is why the formulation of equation
(\ref{eqn_ra}) is misleading.  Here are two reasons.  First, using
1-dimensional reasoning, in an accelerating wind, the behavior of
$v_{r \rm A}$ mitigates the response of $r_{\rm A}$ to the parameters.
For example, for an increase in $B_*$, the Alfv\'en radius will become
larger, but since the flow is accelerating, $v_{r \rm A}$ will also
increase.  \citet{kawaler88} made the approximation that $v_{r \rm A}$
equals the escape speed at $r_{\rm A}$.  In this case, $v_{r \rm A}$
{\it decreases} with radius, giving the opposite effect of an
accelerating wind.  Similarly, the approximations of \citet{mestel84}
that $v_{r \rm A}$ is constant for a slow rotator and proportional to
$\Omega_* r_{\rm A}$ for a fast rotator do not well-approximate the
acceleration exhibited in the winds we simulated.  The second reason
for the surprisingly weak dependence of $r_{\rm A}$ on parameters is
in the amount of open magnetic flux that participates in the flow,
which again is not included in the derivation of equation
(\ref{eqn_ra}), and which again mitigates the effect of parameters on
$r_{\rm A}$.  For example, for an increase in $B_*$, a smaller area on
the stellar surface will have open flux \citep[e.g., compare figures
  \ref{fig_fiducial} and \ref{fig_b400g}, and see][]{mestelspruit87},
so $r_{\rm A}$ will not increase as much as expected in the magnetic
power law formulation.

In future work, it will be important to extend equation
(\ref{eqn_rasim}) to include the effects of rotation, etc.
Furthermore, much work is needed to explore the full consequences
(e.g., for main sequence stars) of the significantly smaller exponent
we find, compared to many previous works.


\section{Summary and Conclusions} \label{sec_discussion}

Using 2D (axisymmetric) MHD simulations, we computed steady-state,
stellar wind solutions for a parameter range appropriate for T Tauri
stars.  We carried out a parameter study including variations of the
stellar mass, radius, surface magnetic field strength, and rotation
rate, as well as mass loss rate, wind acceleration rate, and two
different magnetic geometries (dipole and quadrupole).  Our solutions
enabled us to determine the angular momentum carried in the wind, and
its dependence on many of the parameters of the system.  Our main
conclusions can be summarized as follows:

\begin{enumerate}

\item{For fiducial parameters, the torque is of the same order ($\sim
10^{36}$ erg) as estimated in Paper I.  Therefore, if the stellar
winds of TTSs have similar parameters to those considered here, they
should have a significant influence on the stellar spin.}

\item{The stellar winds are in the regime of moderately fast magnetic
rotator winds.  They produce jets, as well as a wide-angle flow
\citep[see, e.g.,][]{mattbalick04}, which should interact with, and be
modified by, surrounding material \citep[not included in our
simulations; e.g.,][]{gardinerfrank03, shangea06}.}

\item{The cases with quadrupole fields resulted in a torque that is
much weaker than cases with a dipole field of the same surface field
strength.  Specifically, we find that a 200 G dipole field exerts the
same stellar wind torque upon a star as a 1--2 kG quadrupole.  This
illustrates the very strong effect of magnetic geometry on the stellar
wind torque.}

\item{We ran cases where the mass loss rate and other parameters were
fixed, but the thermal wind driving parameters were varied.  For large
variations in the wind acceleration, the torque changed by less than a
factor of 2.  This suggests that the details of the velocity profile
are not of fundamental importance, and our solutions should be a
reasonable approximation for winds with other wind driving mechanisms.
However, it will still be important for future work to compare our
torque results to stellar wind solutions that use alternative driving
mechanisms.}

\item{Our determination of the torque allowed us to calculate the
  Alfv\'en radius (via eq.\ \ref{eqn_tw}), which is a fundamental
  quantity in MHD wind theory.  We compared our numerical solutions to
  previous analytic work and obtained a semi-analytic formulation for
  $r_{\rm A} / R_* \propto [B_*^2 R_*^2 / (\dot M_{\rm w} v_{\rm
      esc})]^m$, with $m \approx 0.22$ (eq.\ \ref{eqn_rasim}), that
  well-describes many of our simulations with dipole fields.  This
  formulation appears to be an improvement over existing work, and the
  exponent $m$ is significantly smaller than usually assumed.}

\end{enumerate}

We will continue to develop the theory of accretion-powered stellar
winds in forthcoming work.  In a companion paper (the third in our
series), we compare the stellar wind torques computed here to the
torques expected to arise from the interaction between the star and an
accretion disk.  We find spin-equilibrium (net zero torque) solutions
and test the suggestion of Paper~I.  In a later paper, we will use the
stellar wind solutions of this work to compute emission properties of
TTS coronal winds.


\acknowledgements

Initial results of various aspects of this work were presented at a
number of meetings, and we wish to thank many participants who
provided interesting questions, discussion, and ideas including: Gibor
Basri, Sylvie Cabrit, Steve Cranmer, Andrea Dupree, Suzan Edwards,
Christian Fendt, Will Fischer, Shu-ichiro Inutsuka, Chris Johns-Krull,
Marina Romanova, Frank Shu, Keivan Stassun, Asif ud-Doula, Jeff
Valenti, and others.  We also thank the referee, Ruben Krasnopolsky,
for his useful suggestions for improving the paper.  SM is supported
by the University of Virginia through a Levinson/VITA Fellowship
partially funded by The Frank Levinson Family Foundation through the
Peninsula Community Foundation.  REP is supported by a grant from
NSERC.






\label{lastpage}
\end{document}